\documentclass[useAMS,usenatbib]{mn2e}
\usepackage{graphicx}

\title[Constraining cosmologies with fundamental constants I. Quintessence and K-Essence]
{Constraining cosmologies with fundamental constants I. Quintessence and K-Essence}
\author[Rodger I. Thompson, C.J.A.P. Martins and P.E. Vielzeuf]{Rodger I. Thompson$^{1}$\thanks{E-mail:
rit@email.arizona.edu (RIT);  Carlos.Martins@astro.up.pt (CJAPM); up110370652@alunos.fc.up.pt 
(PEV)}, C.J.A.P. Martins$^{2,3}$ and P.E. Vielzeuf$^{2,4,5,}$\\
$^{1}$Steward Observatory, University of Arizona, Tucson, AZ 85721, USA\\
$^{2}$Centro de Astrof\'{i}sica, Universidade do Porto, Rua das Estrelas, 4150-762 Porto, Portuagal\\
$^{3}$Department of Applied Mathematica and Theoretical Physics, Centre for Mathematical Sciences,\\
University of Cambridge, Wilberforce Road, Cambridge CB3 OWA, United Kingdom\\
$^{4}$Faculdade de Ciencias, Univeridade do Porto, Rua do Campo Alegre 687, 4169-007 Porto, Portuagal\\
$^{5}$Universit\`{e} Paul Sabatier--Toulouse III, 118 Route de Narbonne 31062 Toulouse Cedex 9, France}
\begin{document}

\date{Accepted xxxx. Received xxxx; in original form xxxx}

\pagerange{\pageref{firstpage}--\pageref{lastpage}} \pubyear{2011}

\maketitle

\label{firstpage}

\begin{abstract}

Many cosmological models invoke rolling scalar fields to account for the observed
acceleration of the expansion of the universe.  These theories generally
include a potential $V(\phi)$ which is a function of the scalar field $\phi$.
Although $V(\phi)$ can be represented by a very diverse set of functions, 
recent work has shown the under some conditions, such as the slow roll
conditions, the equation of state parameter $w$ is either independent
of the form of $V(\phi)$ or is part of family of solutions with only a
few parameters.  In realistic models of this type the scalar field couples to 
other sectors of the model leading to possibly observable changes in the fundamental
constants such as the fine structure constant $\alpha$ and the proton to
electron mass ratio $\mu$.  Although the current situation on a possible
variance of $\alpha$ is complicated there are firm limitations on
the variance of $\mu$ in the early universe.  This paper explores the limits
this puts on the validity of various cosmologies that invoke rolling scalar
fields.  We find that the limit on the variation of $\mu$ puts significant
constraints on the product of a cosmological parameter $w+1$ times a new
physics parameter $\zeta_{\mu}^2$, the coupling constant between $\mu$ and
the rolling scalar field. Even when the cosmologies are restricted to
very slow roll conditions either the value of $\zeta_{\mu}$ must be at 
the lower end of or less than its expected values or the value of $w+1$ must
be restricted to values vanishingly close to 0. This implies that either the
rolling scalar field is very weakly coupled with the electromagnetic field,
small $\zeta_{\mu}$, very weakly coupled with gravity, $(w+1) \approx 0$ or
both. These results stress that adherence to the measured invariance in $\mu$ 
is a very significant test of the  validity of any proposed cosmology and any 
new physics it requires. The limits  on the variation of $\mu$ also produces a 
significant tension with the reported changes in the value of $\alpha$.

\end{abstract}


\section{Introduction} \label{s-intro}  
Tracing the values of the fundamental constants through the history of the universe 
provides strong constraints on the possibility of cosmologies other than the 
standard $\Lambda$CDM universe and new physics that deviates from the standard model.
In this investigation we use the observed limits on the variation of the proton to
electron mass ratio $\mu$ as a new input parameter for three quintessence cosmologies
and K-Essence.  Each of  these cosmologies postulates a rolling scalar field $\phi$ 
with a potential $V(\phi)$.  Realistic models of this class expect the scalar field
to also have non-zero couplings to sectors other than gravity unless an unknown symmetry is 
postulated to suppress them \citep{car98}. Here we assume the simplest non-vanishing
coupling to the electromagnetic sector and a unification scenario of the type described
in \citet{coc07} that leads to a change in $\mu$ that is related to a change in $\alpha$. 

At the epoch of 
each $\frac{\Delta \mu}{\mu}$ measurement there is a constraint placed on the product 
of the coupling of $\mu$ to the rolling scalar field and the equation of state parameter 
$w$ that is independent of the cosmology except for the form of the equation describing 
the dark energy density $\Omega_{\phi}(z)$. The evolution of $\mu$ and $w$ within those 
constraints, however, is dependent on the particular cosmology. We investigate one freezing 
cosmology, slow roll quintessence, and 3 thawing cosmologies, 
hilltop quintessence, non-minimal quintessence and K-Essence. Freezing models start 
with the equation of state parameter w different from $-1$ at early times and 
approaching $-1$ at the present time while thawing models start with w close to 
$-1$ at early times and deviate from $-1$ at the present epoch. Each of these 
cosmologies has the advantage of having a family of solutions which is a function 
of a small number of parameters (\citet{dut11},\citet{gup09},\citet{chi09},\citet{dut08})
We will follow the methodology laid out in \citet{thm12} for just slow roll 
quintessence to investigate the constraints on all four cosmologies.

\section{Observational Constraints} \label{s-obcon}
\begin{figure}
  \vspace{150pt}
\resizebox{\textwidth}{!}{\includegraphics[0in,0in][14.in,3.in]{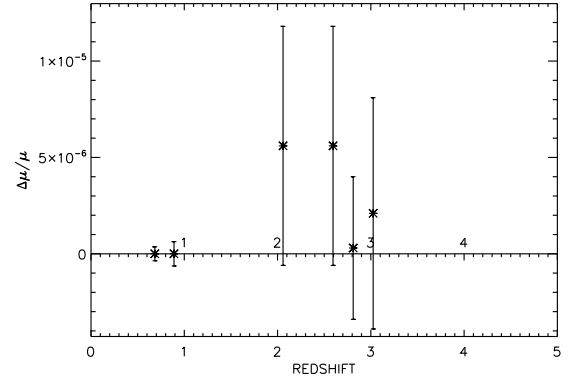}}
  \caption{The observed values of $\Delta \mu / \mu$ and their associated errors
from Table~\ref{tab-con}.  Note that the two lowest redshift errors are $3\sigma$
errors while the rest are $1\sigma$ error bars.} \label{fig-err}
\end{figure}

Appendix~\ref{s-curs} gives a comprehensive list of measurements of $\mu$ using 
astronomical observations. Based on this list Table~\ref{tab-con} gives the
most constraining limits on the value of $\Delta \mu / \mu$.  The listings for
radio observations of PKS1830-211 and B0218+357 give the $3\sigma$ limits about
a null result.  All of the other observations contain the null result in their
$1\sigma$ limits. Figure~\ref{fig-err} shows the errors from Table~\ref{tab-con} 
plotted as a function of redshift.  These are the measurements used in establishing 
the constraints used in this analysis.

\begin{table}
 \begin{minipage}{120mm}
\begin{tabular}{lllllll}
\hline
Object & Redshift  & $\Delta \mu / \mu$ & error & $(w+1) \zeta_{\mu}^2$ & Accuracy & Reference\\
\hline
Q0347-383 & $3.0249$ & $2.1 \times 10^{-6}$ & $\pm 6. \times 10^{-6}$&$\leq 3.8 \times 10^{-11}$ & $1\sigma$ & \citet{wen08}\\
Q0405-443 & $2.5974$ & $10.1 \times 10^{-6}$ & $\pm 6.2 \times 10^{-6}$&$\leq 4.0 \times 10^{-11}$ & $1\sigma$ & \citet{kin09}\\
Q0528-250 & $2.811$ & $3.0 \times 10^{-7}$ & $\pm 3.7 \times 10^{-6}$&$\leq 1.4 \times 10^{-11}$ & $1\sigma$ & \citet{kin11}\\ 
J2123-005 & $2.059$ & $5.6 \times 10^{-6}$ & $\pm 6.2 \times 10^{-6}$&$\leq 4.0 \times 10^{-11}$ & $1\sigma$ & \citet{mal10}\\
PKS1830-211 & $0.89$ & $0.0$ & $\pm 6.3 \times 10^{-7}$&$\leq 6.5 \times 10^{-13}$ & $3\sigma$ & \citet{ell12}\\
B0218+357 & $0.6847$ & $0.0$ & $\pm 3.6 \times 10^{-7}$&$\leq 2.8 \times 10^{-13}$ & $3\sigma$ & \citet{kan11}\\
\hline
\end{tabular}
\caption{Observational constraints used in this analysis.}  \label{tab-con}
\end{minipage}
\end{table}

\section{Varying $\mu$ in the Context of New Physics}

A time varying value of $\mu$ is not allowed in the Standard Model so any variation
in $\mu$ introduces new physics.  As in \citet{thm12} we follow the discussion
of \citet{nun04}, hereinafter NL, which actually discusses varying values of the 
fine structure constant $\alpha$.  The same physics applies to $\mu$ with the two
constants connected by
\begin{equation}  \label{eq-amu}
\frac{\dot{\mu}}{\mu} \sim \frac{\dot{\Lambda}_{QCD}}{\Lambda_{QCD}} -\frac{\dot{\nu}}
{\nu} \sim R \frac{\dot{\alpha}}{\alpha}
\end{equation}
given in \citet{ave06}. In equation~\ref{eq-amu} $\Lambda_{QCD}$ is the QCD scale, $\nu$ is 
the Higgs vacuum expectation value and R is a scalar often considered to be on the order
of -40 to -50 \citep{ave06}. In the rest of the discussion we assume the value 
of R to be -40 but consider possible variations from this value in a later 
section on the tension between the limits on the variation of $\mu$ and the reported 
variation of $\alpha$.  NL consider the simplest possible coupling of $\mu$ with a
rolling scalar field $\phi$, namely a linear coupling given by
\begin{equation} \label{eq-dmu}
\frac{\Delta \mu}{\mu} = R \zeta_{\alpha} \kappa (\phi - \phi_0) = \zeta_{\mu} \kappa (\phi - \phi_0)
\end{equation}
where $\zeta_x$ ($x = \alpha, \mu$) is the coupling constant, $\kappa = \frac{\sqrt{8 \pi}}{m_p}$ 
and $m_p$ is the Planck mass. The coupling constants $\zeta_x$ are considered constant in time.
Certainly other forms of coupling can be considered but in the absence of any information on 
its nature we choose to use the simplest form. Further, since it is known by observation that any
variation of $\mu$ is small, a linear coupling approximation is legitimate at least out to redshifts
on the order of $4$. The rolling potential $V(\phi)$ is a function of the scalar
$\phi$  and the equation of state $w$ is given by
\begin{equation} \label{eq-w}
w \equiv \frac{p_{\phi}}{\rho_{\phi}} = \frac{\dot{\phi}^2 - 2 V(\phi)}{\dot{\phi}^2 + 2 V(\phi)}
\end{equation}
(NL) then show that $w +1$ is also given by
\begin{equation} \label{eq-wa}
w + 1 = \frac{(\kappa \phi')^2}{3 \Omega_{\phi}}
\end{equation}
where $\Omega_{\phi}$ is the dark energy density. Here $\dot{\phi}$ and $\phi'$ indicate 
differentiation with respect to cosmic time and to $N = \log{a}$ respectively where $a$ is 
the scale factor of the universe. Equation \ref{eq-dmu} shows that
\begin{equation} \label{eq-phip}
\phi' = \frac{\mu'}{\kappa \zeta_{\mu} \mu}
\end{equation}
It follows from \ref{eq-wa} and \ref{eq-phip} that
\begin{equation} \label{eq-wmu}
w +1 = \frac{(\mu'/\mu)^2}{3 \zeta_{\mu}^2 \Omega_{\phi}} = \frac{(\alpha'/\alpha)^2}{3 \zeta_{\alpha}^2 \Omega_{\phi}}
\end{equation}
which establishes a connection between the evolution of $w$ and $\mu$.  Note that for the phantom
case, $w < -1$, the two right hand terms in equation~\ref{eq-wmu} are preceded by a minus sign.

Since $\mu'=a(\frac{d\mu}{d a})$ we can find the variance of $\mu$ relative to its present
day value at any scale factor $a$ by
performing the integral

\begin{equation} \label{eq-muint}
\frac{\Delta\mu}{\mu} =\zeta_{\mu}\int^{a}_{1}\sqrt{3\Omega_{\phi}(x)(w(x)+1)}x^{-1}dx
\end{equation}
The value of $w+1$ versus redshift or scale factor is set by the different cosmologies.
For phantom cosmologies the factor of $(w+1)$ in equation~\ref{eq-muint} is replaced by
$-(w+1)$.

\section[]{Constraints that are Relatively Independent of the Cosmological Model} \label{s-indep}

\begin{figure}
  \vspace*{75pt}
\resizebox{\textwidth}{!}{\includegraphics[0in,0in][14in,3.in]{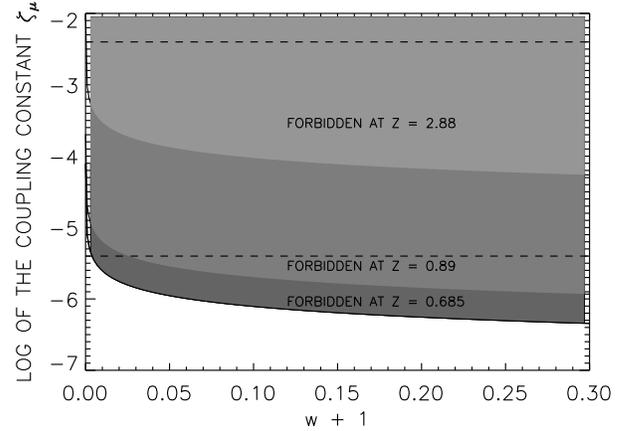}}
  \caption{The figure shows the forbidden and allowed parameter space in the $\zeta_{\mu}$,
$(w+1)$ plane based on the three most restrictive low and high redshift
observations. The upper light shaded area is for the constraint at a redshift of 2.811,
the middle darker area and above are for a redshift of 0.89, and lower dark shaded area
and above is for the constraint at a redshift of 0.685.   The dashed lines indicate the 
upper and lower most likely limits on the coupling factor $\zeta_{\mu}$ as discussed in 
the text.} \label{fig-zmulim}
\end{figure}

Equation~\ref{eq-wmu} provides a constraint on the combination of a cosmological parameter 
$w$ and a new physics parameter $\zeta_{\mu}$ relative to the limits on $\mu'/\mu$ 
\begin{equation} \label{eq-lim}
(w+1)\zeta_{\mu}^2 = \frac{(\mu'/\mu)^2}{3 \Omega_{\phi}}
\end{equation}
that is independent of the form of the potential $V(\phi)$. Again utilizing that
$\mu' = a(\frac{d \mu}{d a})$ we can write
\begin{equation} \label{eq-del}
(w+1)\zeta_{\mu}^2 = \frac{(\Delta \mu/\mu)^2(a/\Delta a)^2}{3 \Omega_{\phi}}
\end{equation}
giving the constraint on $(w+1)\zeta_{\mu}^2$ as a function of $\Delta\mu$ and the
dark energy density $\Omega_{\phi}$. Any combination of a given cosmology and  
value of $\zeta_{\mu}$ must satisfy the constraint given by equation~\ref{eq-del} at the
redshift of the observation.  Different cosmologies, however, take separate paths
in the $\Delta\mu$ redshift plane to meet the constraints.  In order to proceed we
now impose the condition that the dark energy density factor $\Omega_{\phi}$ is given 
by. 
\begin{equation}\label{eq-omega}
\Omega_{\phi} = [1+(\Omega_{\phi 0}^{-1} - 1)a^{-3}]^{-1}
\end{equation}
In equation~\ref{eq-omega} the subscript $0$ refers to the present day value. This form
for $\Omega_{\phi}$ assumes that $w$ is close to $-1$ so that the $e^{-3 \int \frac{(1+w(z))dz}{1+z}}$
term that multiplies $a^{-3}$ in the full expression is approximately 1. This applies 
for the cases considered in this work. An
examination of the exact dark energy density solutions for each of the cosmologies
indicates that most variations from equation~\ref{eq-omega} are less than $10\%$
at redshfits less than 4 with the maximum being $20\%$ for some K-Essence cases.
Figure~\ref{fig-zmulim} is therefore a good representation of the forbidden 
parameter space. The bounds on $(w+1)\zeta_{\mu}^2$ at each epoch are listed in 
Table~\ref{tab-con}.

Another way to look at the constraints imposed by the $\Delta \mu$ limits is to look
at the allowed and forbidden areas in the $\zeta_{\mu}$ $(w+1)$ plane as a function
of redshift. Figure~\ref{fig-zmulim} shows the allowed and forbidden areas for the
most restrictive low redshift constraints, B0218+357 at z = 0.6847 \citep{kan11} 
and PKS 1830-211 at z= 0.89 \citep{ell12} as well as the most restrictive high 
redshift constraint, Q0528-250 at z = 2.811 \citep{kin11}.  In the figure all
of the space above the solid lines is forbidden. Figure~\ref{fig-zmulim} is a
fundamental result of this paper.  It applies to all cosmologies for which
equation~\ref{eq-omega} for the dark energy density is valid.  First shown
in a slightly different format in \citet{thm12}, it defines the allowed parameter
space for $w$ and $\zeta_{\mu}$. All cosmologies must adhere to the allowed
space at the redshifts of the observations.  Different cosmologies take different
paths through the allowed parameter space, therefore, filling in the diagram
with measurements at a large number of redshifts with greatly improved accuracy
is an important task.

\begin{figure}
  \vspace*{75pt}
\resizebox{\textwidth}{!}{\includegraphics[0in,0in][14in,3.in]{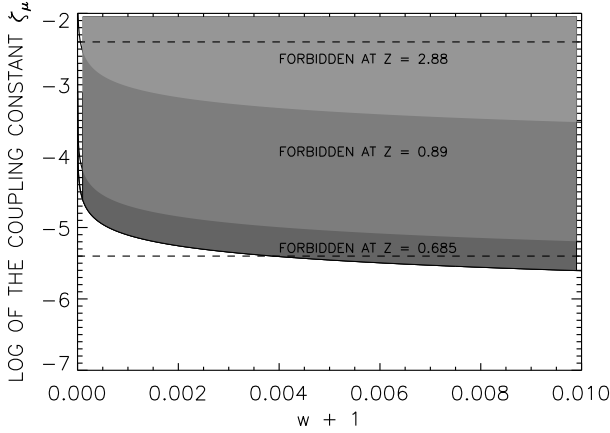}}
  \caption{The figure gives a detailed view of the narrow allowed limits on $w+1$ if the
value of $\zeta_{\mu}$ is taken at its lower expected limit shown by the lower dashed
line. At a redshift of 0.685 $w+1$ is constrained to be less than $0.004$ unless the
coupling constant is reduced below its lowest expected value.} \label{fig-zmuminlim}
\end{figure}

NL use the work of \citet{cop04} to set likely bounds on the value of $\zeta_{\alpha}$
of $\zeta_{\alpha} \sim 10^{-7} - 10^{-4}$.  For a R value of -40 this translates to
a likely range for $\zeta_{\mu}$ of $\zeta_{\mu} \sim -4 \times 10^{-6}$ to 
$- 4 \times 10^{-3}$. These bounds are shown by the dashed horizontal lines 
figure~\ref{fig-zmulim}. Figure~\ref{fig-zmuminlim} is a greatly magnified view
of the $w+1$ space from $0$ - $0.01$ which shows that at the lowest expected value 
of $\zeta_{\mu}$ only the space
with $(w+1) < 0.004$ is allowed at a redshift of 0.685. Coupling constants near 
the high end of the expected value require $(w+1)$ to be essentially zero. However, as
discussed later, setting $\zeta_{\mu}$ to less than $5 \times 10^{-7}$ allows a 
full range of $(w+1)$ values. Inclusion of the two radio observations at redshifts
of 0.685 and 0.89 results in a much more restricted parameter space than presented
in \citet{thm12} that only included the results from optical observations of H$_2$.
We next investigate how figure~\ref{fig-zmulim} impacts the the allowed parameters
for the four cosmologies.

\section{The Parameterized Solutions} \label{s-ps}
Each of the four cosmologies, examined in this work have parameterized solutions
for the value of $w+1$ as a function of scale factor or redshift.  As shown in \citet{thm12}
this also leads to solutions for $\Delta\mu/\mu$ through Equation~\ref{eq-wmu}. In
this section we examine the parameterized solutions for each of the four cosmologies
and the subsequent solutions for the variance of $\mu$.  In each case we use
Geometrized units where $8 \pi G=1$. Once the parameterized solutions are established
reasonable parameters are selected to provide test cases for each cosmology. In 
section~\ref{s-fit} the value of $\zeta_{\mu}$ is then adjusted to satisfy the
constraints on $\Delta \mu$ given in Table~\ref{tab-con}.

\subsection{Slow Roll Conditions}

In each of these cosmologies, except for hilltop quintessence, we impose the standard 
slow roll conditions on the potential $V(\phi)$.

\begin{equation}\label{eq-sr1}
\lambda^2 \equiv (\frac{1}{V} \frac{dV}{d\phi})^2 \ll 1
\end{equation}

\begin{equation}\label{eq-sr2}
|\frac{1}{V} \frac{d^2V}{d\phi^2}| \ll 1
\end{equation}

These conditions produce a very flat potential and are generally the same conditions
for a minimal variation in $\mu$.  This means that the restrictions on the parameter
space for non-slow roll cosmologies would probably be even stricter than in the slow
roll case.  In many cases, such as slow roll quintessence the value of $\lambda$ in
Equation~\ref{eq-sr1} is taken to be a constant value equal to $\lambda_0$ which 
then becomes one of the parameters.

\subsubsection{Slow Roll Quintessence}

This cosmology was already treated in \citet{thm12} but we include it here for completeness.
The dynamical equation is given by
\begin{equation} \label{eq-quin}
\ddot{\phi} + 3H \dot{\phi} + \frac{dV}{d\phi} = 0
\end{equation}
and the parameterized solution for $w+1$ is given by \citet{dut11} as
\begin{equation} \label{eq-qw}
1 + w = \frac{1}{3} \lambda_0^2[\frac{1}{\sqrt{\Omega_{\phi}}}-(\frac{1}{\Omega_{\phi}}-1)
(\tanh^{-1}(\sqrt{\Omega_{\phi}}) + C)]^2
\end{equation} 
The parameter $C$ characterizes the family of solutions and is set by and early condition
on $w =w_i$.
\begin{equation} \label{eq-c}
C = \pm \frac{\sqrt{3(1+w_i)}\Omega_{\phi_i}}{\lambda_0}
\end{equation}
$C$ is set by picking the value of $w$ at some early epoch such as $z=5$ as we will do
in a later section.  This value $w_i$ then sets the solution for $w+1$ and $\Delta\mu/\mu$
at all other epochs.  As shown in \citet{thm12} equation~\ref{eq-wmu} and equation~\ref{eq-omega}
give the evolution of $\mu$ as
\begin{eqnarray} \label{eq-deltamu}
\frac{\Delta \mu}{\mu} = \zeta_{\mu} \lambda_0 \int_{1}^{a} \{1-[(1+(\Omega_{0}^{-1}-1)x^{-3})^{-1/2} \nonumber \\
-(1+(\Omega_{0}^{-1}-1)x^{-3})^{1/2}] \nonumber \\
\times [\tanh^{-1} (1+(\Omega_{0}^{-1}-1)x^{-3})^{1/2}+C]\}x^{-1}dx
\end{eqnarray}

\subsubsection{Hilltop Quintessence}

The dynamical equation for hilltop quintessence is the same as for slow roll quintessence.
In hilltop quintessence the scalar field is rolling down a potential from a position very 
near the maximum of the potential.  The cosmology adheres to the first slow roll condition 
but in some cases the second slow roll condition is relaxed.  In this section we follow the 
discussion of \citet{dut08} in developing the parameterized solutions.  \citet{dut08}
show that $w+1$ is given by
\begin{eqnarray} \label{eq-hqw}
1+w(a) = (1+w_0)a^{3(K-1)} \nonumber \\ 
\frac{[(F(a)+1)^K(K-F(a))+(F(a)-1)^K(K+F(a))]^2}{[(\Omega_{\phi 0}^{-\frac{1}{2}}+1)^K(K-\Omega_{\phi 0}^{-\frac{1}{2}})+(\Omega_{\phi 0}^{-\frac{1}{2}}-1)^K(K+\Omega_{\phi 0}^{-\frac{1}{2}})]^2}
\end{eqnarray} 
where $F(a)$ is given by
\begin{equation}\label{eq-hf}
F(a)=\sqrt{1+(\Omega_{\phi 0}^{-1}-1)a^{-3}}
\end{equation}
The parameter $K$ is given by
\begin{equation} \label{eq-htk}
K=\sqrt{1-(4/3)V''(\phi_*)/V(\phi_*)}
\end{equation}
where $\phi_*$ is the value of $\phi$ at the maximum.  At the maximum $V''(\phi_*)<0$
therefore $K>1$. For true slow roll conditions $K$ should not be much greater than $1$.

The variance of $\mu$ is then given by
\begin{eqnarray} \label{eq-deltahqmu}
\frac{\Delta \mu}{\mu} = \zeta_{\mu} \sqrt{(1+w_0)} \int_{1}^{a}(x^{\frac{3(K-1)}{2}} \frac{\sqrt{3}}{\sqrt{(1+(\Omega_{0}^{-1}-1)x^{-3}}}\nonumber \\
\frac{[(F(x)+1)^K(K-F(x))+(F(x)-1)^K(K+F(x))]}{[(\Omega_{\phi 0}^{-\frac{1}{2}}+1)^K(K-\Omega_{\phi 0}^{-\frac{1}{2}})+(\Omega_{\phi 0}^{-\frac{1}{2}}-1)^K(K+\Omega_{\phi 0}^{-\frac{1}{2}})]})x^{-1}dx
\end{eqnarray}

\subsubsection{Non-Minimal Quintessence and Phantom}
As the name implies, in non-minimal quintessence and phantom cosmologies the quintessence
and phantom fields couple with gravity in a non-minimal way.  We will follow the discussion
of \citet{gup09} who introduce the usual parameter $\epsilon$ which has a value of $+1$
for quintessence and $-1$ for phantom where the value of $w$ is less than $-1$. The dynamical
equation for non-minimal models is given by
\begin{equation} \label{eq-nmdy}
\ddot{\phi}+3H\dot{\phi}+6\xi (\dot{H}+2H^2)\phi+\epsilon V'(\phi)=0
\end{equation}
In equation~\ref{eq-nmdy} $\xi$ is the non-minimal coupling parameter, usually set to $1/6$,
which we will use here.  The results are relatively insensitive to the value, however.
\citet{gup09} show that the parameterized solution for the equation of state is given by
\begin{eqnarray} \label{eq-nmqw}
1+w_{\phi}(a)= \epsilon \frac{1}{9}\{\frac{[1+(\Omega_{\phi 0}^{-1}-1)a^{-3}](1-\Omega_{\phi 0})} {1 + (a^3-1)\Omega_{\phi 0}}\}^{2-(8\xi/3)} \nonumber \\ 
\times \{6 \epsilon \sqrt{2} z_0 \xi B([1+(\Omega_{\phi 0}^{-1}-1)a^{-3}]^{-1};\frac{1}{2}-\frac{4 \xi}{3},-1+\frac{4 \xi}{3}) \nonumber \\
+[\sqrt{3} \lambda_0 (1-2 \xi)-6\epsilon \sqrt{2} z_0 \xi] \nonumber \\
\times B([1+(\Omega_{\phi 0}^{-1}-1)a^{-3}]^{-1};\frac{3}{2}-\frac{4 \xi}{3},-1+\frac{4 \xi}{3})\}^2
\end{eqnarray}
where B is the incomplete Beta function. $\lambda_0$ is the value of the first slow
roll condition and again assumed to be constant. $z_0$ is the average value of the
auxiliary variable $z=\frac{\kappa \phi}{\sqrt{6}}$. Here $\phi$ is the scalar field
and $\kappa^2=8\pi G$.  The nominal value of $z_0$ is $10^{-5}$ with the result
again very insensitive to the value.

Equation~\ref{eq-nmmu} then gives the variation of $\mu$ with scale factor $a$ as

\begin{eqnarray} \label{eq-nmmu}
\frac{\Delta \mu}{\mu} = \frac{\zeta_{\mu}\epsilon}{\sqrt{3}}\int_{1}^{a}\frac{1}{\sqrt{1+(\Omega_{\phi 0}-1)x^{-3}}} \nonumber \\ 
\{\frac{[1+(\Omega_{\phi 0}^{-1}-1)x^{-3}](1-\Omega_{\phi 0})} {1 + (x^3-1)\Omega_{\phi 0}}\}^{1-(4\xi/3)} \nonumber \\ 
\times \{6 \epsilon \sqrt{2} z_0 \xi B([1+(\Omega_{\phi 0}^{-1}-1)x^{-3}]^{-1};\frac{1}{2}-\frac{4 \xi}{3},-1+\frac{4 \xi}{3}) \nonumber \\
+[\sqrt{3} \lambda_0 (1-2 \xi)-6\epsilon \sqrt{2} z_0 \xi] \nonumber \\
\times B([1+(\Omega_{\phi 0}^{-1}-1)x^{-3}]^{-1};\frac{3}{2}-\frac{4 \xi}{3},-1+\frac{4 \xi}{3})\}x^{-1}dx
\end{eqnarray}

Note that the flip in sign in the right hand part of equation~\ref{eq-wmu} cancels the
$\epsilon = -1$ leading equation~\ref{eq-nmmu} making the phantom solutions for $\Delta \mu/\mu$
indistinguishable from the quintessence solutions.

\subsubsection{K-Essence} \label{sss-ke}
In this section we follow the development of thawing slow roll k-essence by \citet{chi09}.
K-essence introduces a non-cannonical kinetic term into the Lagrangian $F(X)$ such that the
pressure is given by
\begin{equation} \label{eq-kp}
p(\phi,X)=V(\phi)F(X)
\end{equation}
where $\phi$ and $V(\phi)$ are again the rolling scalar field and the potential of the field.
$X$ is given by
\begin{equation} \label{eq-x}
X= -\nabla^{\mu}\phi\nabla_{\mu}\phi/2
\end{equation}
The K-Essence equation of motion is given by
\begin{equation} \label{eq-kmot}
\ddot{\phi} + 3{c_{s}}^2 H \dot{\phi} + {c_{s}}^2\frac{2XF_X-F}{F_X} \frac{V'}{V} = 0
\end{equation}
where
\begin{equation} \label{eq-kcs}
{c_{s}}^2= \frac{F_X}{2XF_{XX}+F_X}
\end{equation}
$F_X$ and $F_{XX}$ indicate single and double derivatives with respect to $X$ and
$V'$ is the the derivative of $V$ with respect to $\phi$. The slow roll conditions
are the same as given in equations~\ref{eq-sr1} and~\ref{eq-sr2}

\cite{chi09} show that the equation of state for slow roll k-essence can be parametrized in
the following form.
\begin{eqnarray} \label{eq-kwpar}
1+w(a)=(1+w_0)a^{3(K-1)}  \nonumber \\
(\frac{(K-F(a))(F(a)+1)^K+(K-F(a))(F(a)-1)^K}{(K-\Omega_{\phi 0}^{-1/2})(\Omega_{\phi 0}^{-1/2}+1)^K+(K+\Omega_{\phi 0}^{-1/2})(\Omega_{\phi 0}^{-1/2}-1)^K})^2
\end{eqnarray}
where
\begin{equation}\label{eq-kk}
K=\sqrt{1-\frac{4}{3}\frac{V''(\phi_i)}{F_X(0)V(\phi_i)^2}}
\end{equation}
and
\begin{equation}\label{eq-kf}
F(a)=\sqrt{1+(\Omega_{\phi 0}^{-1}-1)a^{-3}}
\end{equation}
where $F(a)$ in equation~\ref{eq-kf} is not $F(X)$. $\phi_i$ is an initial value of
$\phi$ and $\Omega_{\phi 0}$ and $w_0$ are the present day values of $\Omega_{\phi}$ 
and $w$. For K-Essence the equation for $\frac{\Delta \mu}{\mu}$ is
\begin{eqnarray}\label{eq-kdmu}
\frac{\Delta \mu}{\mu}=\zeta_{\mu}\sqrt{1+w_0}\int_{1}^{a}\frac{\sqrt{3}x^{3\frac{K-1}{2}}}{\sqrt{1+(\Omega_{\phi 0}^{-1} - 1)x^{-3}}} \nonumber \\
\frac{(K-F(x))(F(x)+1)^K+(K-F(x))(F(x)-1)^Kx^{-1}}{(K-{\Omega_{\phi 0}}^{-1/2})(\Omega_{\phi 0}^{-1/2}+1)^K+(K+\Omega_{\phi 0}^{-1/2})(\Omega_{\phi 0}^{-1/2}-1)^K}dx
\end{eqnarray}

For phantom solutions the leading term $\sqrt{1+w_0}$ becomes $\sqrt{-(1+w_0)}$ and as
in the phantom non-minimal case the phantom K-Essence solutions are indistinguishable from the
quintessence solutions.

\section{Fitting the Constraints} \label{s-fit}
Now that the parameterized solutions have been presented we can see what the parameters 
need to be in order to satisfy the constraints presented in Figure~\ref{fig-err}.  Even 
with the parameterized solutions there is an infinite number of cosmologies possible. To 
limit the field the solution space needs to be constrained.  We choose to place the constraints
on the allowed values of the equation of state parameter $w$.

\subsection{Case Values for $w$} \label{ss-cw}
The appropriate case values for $w$ will be different for the one freezing cosmology,
slow roll quintessence, than for the three thawing cosmologies.  For the thawing 
cosmologies we choose present epoch values of $w$ of -0.99, -0.95 and -0.9 as 
being consistent with the slow roll conditions.  For cosmologies allowing phantom 
solutions we choose the mirror solutions of -1.01, -1.05 and -1.1 as well. For the
freezing slow roll quintessence we choose values of $w$ at redshift 5 of -0.5, -0.75
and -0.9.  In each cosmology we then adjust the parameters used in Section~\ref{s-ps} 
to achieve the desired initial values of $w$.

The case values for $w$ are meant to span the range appropriate to the slow roll
conditions with values very close to $-1$ to values $0.1$ deviant from $-1$ that
start to strain the slow roll conditions.  The cases for slow roll quintessence
satisfy the conditions for the redshifts of the observations but start to become
deviant at significantly higher redshifts.  Allowing a larger deviation from $-1$
pushes the values of $\zeta_{\mu}$ even lower, consistent with the constraints
in figure~\ref{fig-zmulim}.

\subsection{Setting the Parameters} \label{ss-par}
\begin{figure}
\vspace{100pt}
\resizebox{\textwidth}{!}{\includegraphics[0in,0in][14in,3.in]{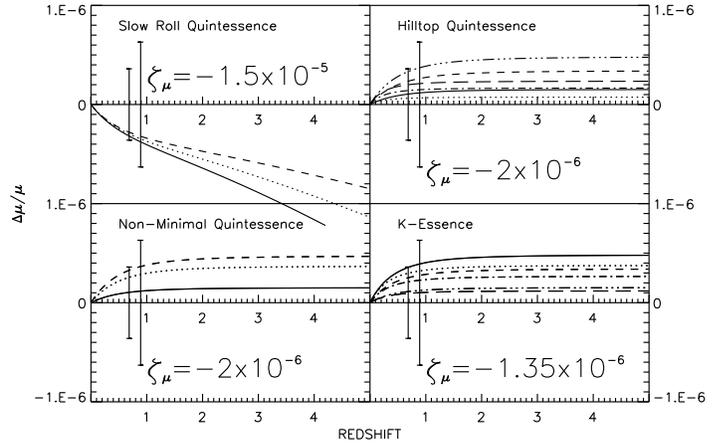}}
  \caption{This figure plots the evolution of $\Delta \mu / \mu$ versus redshift for
each of the four cosmologies. The value of $\zeta_{\mu}$ has been adjusted in each
cosmology so that all cases for that cosmology fall within the observational 
constraints. The higher redshift constraint at z=2.811 is not plotted since it
is larger than the plot size. The value of $\zeta_{\mu}$ is marked in the lower left of each plot.
Refer to Table~\ref{tab-p} for the line style for each case. NB The value of
$\zeta_{\mu}$ in the figure is the value shown in the figure not the values
in Table~\ref{tab-p}.} \label{fig-fit}
\end{figure}

Having chosen either the present day or redshift 5 values of the equation of state 
parameter $w$ for the cosmologies we then vary $\zeta_{\mu}$ to satisfy the 
$\Delta \mu / \mu$ constraints since the $w(a)$ tracks are independent of $\zeta_{\mu}$.
First, in order to show the effect of the parameters on the solutions in figure~\ref{fig-fit},
we choose a single value for $\zeta_{\mu}$ such that all solutions for the parameter 
suite for a given cosmology fit the constraints. In Table~\ref{tab-p}, however, we 
list the largest absolute value of $\zeta_{\mu}$ that fits the constraints for each 
individual parameter set along with the values of the parameters.

In the slow roll quintessence cosmology we only use negative values of the C parameter as
they represent solutions to a field rolling down the potential. Positive values 
represent cases where the field initially rolls uphill \citep{dut11}. From \citet{dut11}
we choose $\lambda_0 = 0.08$ for this cosmology. Note that we could equally well have
chosen to vary $\lambda_0$ instead of $\zeta_{\mu}$ to meet the constraints on slow
roll cosmology, however that would have changed the value of the parameter $C$ in 
equation~\ref{eq-c}. In hilltop quintessence we choose two values of K, $1.01$ and $4.0$ to 
represent a very slowly rolling solution for $1.01$ and a solution, $K=4.$, in which 
the field rolls faster. In non-minimal quintessence we use the nominal values for
$\xi$ and $z_0$ of $1/6$ and $10^{-5}$ given in \citet{gup09}, however, as noted above
the solutions are extremely insensitive to large changes in either of these parameters.
The desired values of $w$ in non-minimal quintessence are achieved by adjusting the slow
roll parameter $\lambda_0$. The phantom solutions are produced by setting $\epsilon$ to
$-1$ instead of $+1$. K-Essence also has phantom solutions.  The K values for K-Essence
are set by Equation~\ref{eq-kk} rather than by Equation~\ref{eq-htk} for the hilltop
quintessence case.  Since the potential is not starting at its maximum value the value
of K can be less than $1$.  We bound the cases by letting $K=0.1,2.0$.

It is clear that even with the limited excursions of $w$ from $-1$ fitting the constraints
requires the absolute values of $\zeta_{\mu}$ in the lower range of expected values and in 
some cases lower than the lowest expected value of $-4 \times 10^{-6}$. Given the softness
of the boundaries this result should probably taken as guidance in further calculations as
opposed to invalidation of the concept.  The results, however, are consistent with the 
Standard Model in which no variation in $\mu$ is expected.

\begin{table*}
 \begin{minipage}{120mm}
\begin{tabular}{llllllll}
\hline
Cosmology & $\zeta_{\mu}$  &  $w$\footnote{$w$ values for slow roll quintessence are for redshift =5, all others are at redshift 0.} &   $C$ & $K$ & $ \lambda_0$ & $(w+1)_{0.685}$\footnote{The value of $w+1$ at a redshift of 0.685}& linestyle\\
\hline
Slow Roll Quintessence & $-1.69\times 10^{-5}$ & $-0.5$  & $-0.163611$ & - & $0.08$ & $0.0012$& solid \\
                    & $-1.81\times 10^{-5}$ & $-0.75$ & $-0.11569$ & - & $0.08$ & $0.00090$ & dotted \\
                    & $-1.94\times 10^{-5}$ & $-0.9$  & $-0.073169$ & - & $0.08$ & $0.00067$ & dash \\
Hilltop Quintessence& $-6.85 \times 10 ^{-6}$ & $-0.99$ & - & $1.01$ & - & $0.00037$ & solid\\
                    & $-1.14 \times 10 ^{-5}$ & $-0.99$ & - & $4.0$ & - & $0.00036$ & dotted \\
                    & $-3.06 \times 10 ^{-6}$ & $-0.95$ & - & $1.01$ & - & $0.018$ & dash\\
                    & $-5.10 \times 10 ^{-6}$ & $-0.95$ & - & $4.0$ & - & $0.0018$ & dash dot\\
                    & $-2.16 \times 10 ^{-6}$ & $-0.9$ & - & $1.01$ & - & $0.037$& dash 3dot\\
                    & $-3.61 \times 10 ^{-6}$ & $-0.9$ & - & $4.0$ & - & $0.0036$ & long dash\\
Non-Minimal Quintessence & $-6.88 \times 10 ^{-6}$ & $-0.99$ & - &  - & $0.32$ & $0.0036$& solid \\
                    & $-2.81 \times 10 ^{-6}$ & $-0.95$ & - & - & $0.782$ & $0.021$ & dotted \\
                    & $-2.20 \times 10 ^{-6}$ & $-0.9$ & - & - & $1.0$ & $0.035$ & dash\\
                    & $6.88 \times 10 ^{-6}$ & $-1.01$ & - &  - & $0.32$ & $-0.0036$ & solid \\
                    & $2.81 \times 10 ^{-6}$ & $-1.05$ & - & - & $0.782$ & $-0.021$ & dotted\\
                    & $2.20 \times 10 ^{-6}$ & $-1.1$ & - & - & $1.0$ & $-0.035$ & dash \\
K-Essence           & $-1.36 \times 10 ^{-6}$ & $-1.1$ & - & $0.1$ & - & $-0.045$ & solid\\
                    & $-1.63 \times 10 ^{-6}$ & $-1.1$ & - & $2.0$ & - & $-0.021$ & dotted\\
                    & $-1.93 \times 10 ^{-6}$ & $-1.05$ & - & $0.1$ & -& $-0.023$& dash \\
                    & $-2.31 \times 10 ^{-6}$ & $-1.05$ & - & $2.0$ & - & $-0.010$& dash dot\\
                    & $-4.31 \times 10 ^{-6}$ & $-1.01$ & - & $0.1$ & - & $-0.0046$& dash 3dot\\
                    & $-5.16 \times 10 ^{-6}$ & $-1.01$ & - & $2.0$ & - & $-0.00021$& long dash\\
                    & $-4.31 \times 10 ^{-6}$ & $-0.99$ & - & $0.1$ & - & $0.0045$ & long dash\\
                    & $-5.16 \times 10 ^{-6}$ & $-0.99$ & - & $2.0$ & - & $0.0021$ &dash 3dot\\
                    & $-1.93 \times 10 ^{-6}$ & $-0.95$ & - & $0.1$ & - & $ 0.023$ & dash dot\\
                    & $-2.31 \times 10 ^{-6}$ & $-0.95$ & - & $2.0$ & - & $0.010$ & dash\\
                    & $-1.36 \times 10 ^{-6}$ & $-0.90$ & - & $0.1$ & - & $0.045$ & dotted\\
                    & $-1.63 \times 10 ^{-6}$ & $-0.90$ & - & $2.0$ & - & $0.021$ & solid\\
\hline
\end{tabular}
\caption{Observational constraints used in this analysis. The last column labeled linestyle
indicates the linestyle used for that case in figure~\ref{fig-fit}.}  \label{tab-p}
\end{minipage}
\end{table*}

\subsubsection{The Evolution of $w+1$}

\begin{figure}
  \vspace*{75pt}
\resizebox{\textwidth}{!}{\includegraphics[0in,0in][14in,3.in]{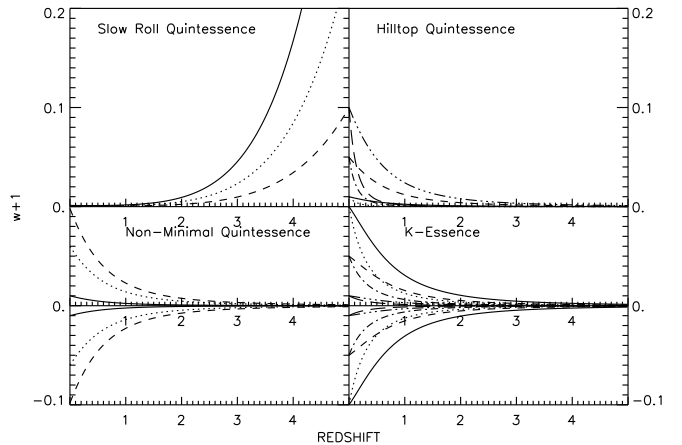}}
  \caption{The figure shows the evolution of the equation of state parameter $w$
by plotting the value of $w+1$ as a function of redshift for each of the four
cosmologies. The last column of Table~\ref{tab-p} contains the line
style code for each of the cases.} \label{fig-w}
\end{figure}

Although each cosmology satisfies the same constraints the evolution of the equation
of state $w$ differs significantly. In particular the freezing slow roll quintessence
cosmology can have $w$ values significantly different than $-1$ at early times and
still satisfy the most restrictive constraint at redshift $0.6847$. Figure~\ref{fig-w}
shows the evolution of the value of $w+1$ for each of the four cosmologies using the
parameters shown in Table~\ref{tab-p}. The evolution of $w$ is of course independent
of the value of $\zeta_{\mu}$.

The 7th row of Table~\ref{tab-p} lists the value of $w+1$ at a redshift of 0.6847
for each cosmology solution. All of these values are quite low corresponding to
the allowed $w+1$ space in Figure~\ref{fig-zmulim}. Slow roll quintessence was
able to satisfy the $\Delta \mu / \mu$ constraints with higher values of $\zeta_{\mu}$
than the other cosmologies and therefore has corresponding smaller deviations of
$w+1$ from $0$ as required by figure~\ref{fig-zmulim}. This cosmology predicts 
very little deviation from $w+1 = 0$ out to a redshift of 2, the redshift
region expected to be probed by currently proposed dark energy space missions. 
Cosmologies such as K-Essence that require very low absolute values of $\zeta_{\mu}$ 
are able to achieve more significant deviations of $w+1$ from $0$. Imposing a value 
of $\zeta_{\mu}$ less that $3 \times 10^{-7}$ provides a large range of possible $w$ 
values. If, however, the expected lower limit of $\zeta_{\mu}=-4\times 10^{-6}$
is imposed then the allowed deviation from $w=-1$ at $z=0.6847$ is only about $0.004$.

\section{New Physics Implications} \label{s-npi}

Given the current constraints on $\Delta \mu/\mu$ any significant deviation of $w$
from $-1$ requires a very low value of $\zeta_{\mu}$ and a deviation greater than
$0.004$ requires a $\zeta_{\mu}$ below the expected lower limit. The strong limits
on the variance of $\mu$ at redshifts below 1 require that the coupling of the scalar
field with either or both of the gravitational and electromagnetic fields be very
weak during that epoch.

If we restrict ourselves to the assumptions we have used so far (slow-roll,
a linear coupling, and a fixed value of $R=-40$) then the spectroscopic
bounds on $\mu$ require the field to have a gravitational behavior almost
exactly like that of a cosmological constant, even deep into the matter
era. Given current theoretical expectations, such a behavior requires
considerable fine-tuning (which is some ways is similar to the fine-tuning
required to have a small but non-zero cosmological constant).

It is of course possible that not all of the above assumptions are correct.
Slow-roll is observationally motivated at low redshifts (say $z<0.5$) when
dark energy dominates and the universe is accelerating, but need not hold
at higher redshifts. For example, one could envisage situations where the
field is moving relatively fast deep in the matter era, but then abruptly
freezes at low redshift, perhaps due to a phase transition associated with
the onset of dark energy domination. This scenario was briefly discussed
(for the case of $\alpha$), in \citet{nun09}.

Similarly, an assumption of linear coupling between the scalar field and
the electromagnetic sector is a reasonable approximation at low redshifts
but can conceivably break down at higher redshifts. The $R$ parameter
can have a different value, which points to a unification scenario
that differs from the ones currently considered to be best motivated
(\citet{ave06}, \citet{coc07}). For example, an $R$ of order unity (in
absolute value) suggests a scenario where unification occurs at
relatively low energies, as is typically the case in models with large
extra dimensions.

Finally, the coupling $\zeta_{\mu}$ itself can be much smaller than
anticipated. This degree of freedom is not independent from the others (in
the context of the models being considered): given a certain non-zero level
of $\mu$ variation, a smaller coupling requires a faster moving field,
and at some point the field must be moving so fast that the slow-roll
approximation breaks down for such a field. In the limiting case the
coupling can be exactly zero, and there would be no variation; however,
as explained in \citet{car98} this is again contrary to the
simplest expectations for realistic models, as some unknown symmetry is needed 
to suppress the coupling.

Which of these scenarios is the correct one is not a question that our
results can answer. However, our analysis highlights that
the current results are at odds with our simplest expectations
regarding scalar field models. At a more general level, this also
highlights that null measurements can be extremely useful
in constraining many theoretical scenarios.

\section{Sandage Loeb Test Values for the Fitted Cosmologies}

In the era of large telescopes with the possibility of very high resolution spectrometers
such as PEPSI and CODEX there has been discussion of direct measurements of the redshift
drift due to the change in the expansion rate of the universe over time \citep{loe98}.
This is generally called the Sandage Loeb Test. It has been recently considered as a
method for measuring the dark energy component through the direct measurement of the
drift in the redshift due to the accelerating expansion of the universe \citep{vie12}.
The change in velocity is given by
\begin{equation} \label{eq-slt}
\Delta v = cH_0t(1-\sqrt{(1+z)\Omega_m + \frac{(1+z)^{-2}\Omega_{\phi}}{e^{-3\int_{0}^{z}\frac{w(x)+1}{1+x}dx}}+\Omega_k})
\end{equation}
where $\Omega_m, \Omega_{\phi},$ and $\Omega_k$ are the ratio of the matter density,
dark energy density and curvature density to the critical density and the equation of
state evolution $w(z)$ is dependent on the cosmology.

\begin{figure}
  \vspace*{75pt}
\resizebox{\textwidth}{!}{\includegraphics[0in,0in][14in,3.in]{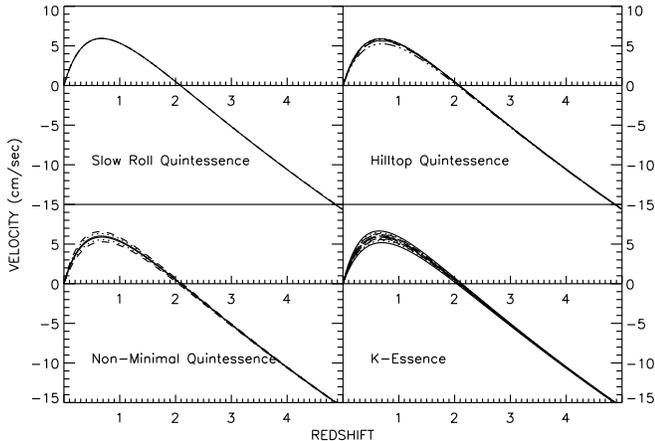}}
  \caption{The figure shows the Sandage Loeb Test velocities as a function of 
redshift for a twenty year baseline.  Where resolvable, the line styles are the 
same as in the previous plots.} \label{fig-slt}
\end{figure}

Figure~\ref{fig-slt} shows the Sandage Loeb Test velocity drifts for a twenty year
baseline.  All of the curves are very close to a $\Lambda$CDM signal, particularly
for the slow roll quintessence cosmology. As pointed out by \citet{vie12}, a larger
coupling factor leads to a slower moving scalar field and less deviation from the
$\Lambda$CDM evolution. In contrast eg. to the results of \citet{bal07}, the lack
of significant dispersion in the Sandage Loeb Test curves is indicative of reduced
parameter space resulting from the $\Delta \mu / \mu$ constraints on the cosmologies
considered here.

\section{Implications on varying $\alpha$} \label{s-alpha}

The observed invariance of $\mu$ appears to be in tension with the reported temporal
and spatial variance of $\alpha$ \citep{kin12}.  Although reported to be a spatial
dipole we consider only the magnitude of the variance which is $\Delta \alpha / \alpha = 
1 \times 10^{-5}$ within the reported errors at an average redshift of 2 for the high
redshift group.  This compares with a conservative bound of $\Delta \mu / \mu < 5 \times 
10^{-6}$ for the same redshift from the observations referenced in this work. If both
results are considered to be correct it requires that the value of $R$ in equation~\ref{eq-amu}
be $0.5$ or less.  This in turn requires that the values of $\frac{\dot{\Lambda}_{QCD}}
{\Lambda_{QCD}}$ and $\frac{\dot{\nu}}{\nu}$ be very similar, contrary to generic GUTS
models \citep{ave06}.  Another possibility is that only the Higgs VEV $\nu$ changes
and the quantum chromodynamic scale $\Lambda_{QCD}$ is constant. Since the Higgs VEV
scales all masses similarly to first order the ratio of the proton to electron mass 
remains unchanged while the Higgs VEV changes in $\alpha$ would be observed. See, however,
\citet{coc07} for a counter argument against varying one parameter and not the other. 
\citet{bar05} present the interesting opposite case of a constant $\alpha$ with a
varying $\mu$. If we entertain the possibility that the reported variation in $\alpha$ 
is erroneous then neither constant has varied, consistent with a $\Lambda$CDM cosmology 
and the Standard Model of physics.

\section{Conclusions} \label{s-con}

No variation in the value of $\mu$ has been found to varying degrees of accuracy at six
different redshifts between 0.685 and 3.02.  This finding is consistent with either or 
both of $\Lambda$CDM cosmology and the Standard Model of Physics being valid. If,
instead, the acceleration of the universe is due to a rolling scalar field that is both
coupled to gravity and the electromagnetic field then one or both of the couplings has to 
be very weak as demonstrated by the very narrow allowed $w+1$  space in figure~\ref{fig-zmulim}
for any significant value of $\zeta_{\mu}$. Slow roll quintessence satisfies the invariance 
of $\mu$ constraints with a low but reasonable $\zeta_{\mu}$ value but with very
minimal values of $w+1$ from the present day out to redshifts of 2.  The invariance of
$\mu$ is in tension with the reported variance of $\alpha$ and requires a ratio of 
$\Lambda_{QCD}$ change to $\nu$ change much closer to $1$ than expected.  Given these
conclusions the value of the fundamental constants as a function of redshift serves
as a powerful constraint on new cosmologies and physics.

\section*{Acknowledgments}
This work has been supported in part by the project PTDC/FIS/111725/2009 (from
FCT, Portugal), and by the joint Master in Astronomy of the Universities
of Porto and Toulouse, supported by project AI/F-11 under the CRUP/Portugal-CUP/France 
cooperation agreement (F-FP02/11). The work of CJM is supported by a Ci\^encia2007 Research 
Contract, funded by FCT/MCTES (Portugal) and POPH/FSE (EC).  RIT gratefully acknowledges
the opportunity for two weeks of visitation at the Centro de Astrof\'{i}sica, Universidade do 
Porto under project PTDC/FIS/111725/2009 (from FCT, Portugal).  RIT also would like to 
acknowledge very helpful discussions with F. Ozel, D. Psaltis, D. Marrone and. J. Bechtold

\appendix

\section{Current determinations of $\Delta \mu/\mu$}
\label{s-curs}
Table~\ref{tab-obs} lists the current determinations of $\Delta \mu/\mu$ in distant
galaxies and in the Milky Way. A subset of the most recent constraints are used in
figures~\ref{fig-err} and~\ref{fig-zmulim}.

\begin{table*} \label{tab-obs}
 \begin{minipage}{120mm}
\begin{tabular}{llll}
\hline
Object & Reference & Redshift & $\Delta\mu/\mu$ \\

\hline
Q0347-383 & \citet{iva02} & 3.0249 & $(5.7 \pm 3.8) \times 10^{-5}$\\
Q0347-383 & \citet{iva02} & 3.0249 & $(12.5 \pm 4.5) \times 10^{-5}$\\
Q0347-383 & \citet{iva03} & 3.0249 & $(\le 8 \times 10^{-5}$\\
Q0347-383 & \citet{uba04} & 3.0249 & $(-0.5 \pm 3.6) \times 10^{-5}$\\
Q0347-383 & \citet{wen08} & 3.0249 & $(2.1 \pm 6) \times 10^{-6}$\\
Q0347-383 & \citet{kin09} & 3.0249 & $(8.2 \pm 7.4) \times 10^{-6}$\\
Q0347-383 & \citet{thm09} & 3.0249 & $(-2.8 \pm 1.6) \times 10^{-5}$\\
Q0347-383 & \citet{wen11} & 3.0249 & $(1.5 \pm 1.1) \times 10^{-5}$\\
Q0347-383 & \citet{wen12} & 3.0249 & $(4.3 \pm 7.2) \times 10^{-6}$\\
347 \& 405 & \citet{iva05} & comb & $(1.64 \pm 0.74) \times 10^{-5}$\\
347 \& 405 & \citet{rei06} & comb & $(2.46 \pm 0.6) \times 10^{-5}$\\
347 \& 405 & \citet{uba07} & comb & $(2.45 \pm 0.59) \times 10^{-5}$\\
Q0405-443 & \citet{thm09} & 2.5974 & $(3.7 \pm 14) \times 10^{-6}$ \\
Q0405-443 & \citet{kin09} & 2.5974 & $(10.1 \pm 6.2) \times 10^{-6}$ \\
Q0528-250 & \citet{fol88} & 2.811 & $\le 2.4\times 10^{-4}$ \\
Q0528-250 & \citet{cow95} & 2.811 & $\le 7.0\times 10^{-4}$ \\
Q0528-250 & \citet{pot98} & 2.811 & $\le 2.0\times 10^{-4}$ \\
Q0528-250 & \citet{kin09} & 2.811 & $(1.4 \pm 3.9)\times 10^{-6}$ \\
Q0528-250 & \citet{kin11} & 2.811 & $(0.3 \pm 3.7)\times 10^{-6}$ \\
J2123-005 & \citet{mal10} & 2.059  & $5.6 \pm 6.2) \times 10^{-6}$\\
PKS 1830-211 &\citet{hen09} & 0.89 & $\le 1. \times 10^{-6}$ \\
PKS 1830-211 &\citet{mul11} & 0.89 & $\le 2. \times 10^{-6}$ \\
PKS 1830-211 &\citet{ell12} & 0.89 & $(\le 6.3 \times 10^{-7}$ \\
B0218+357 & \citet{fla07} & 0.6847 & $(0.6 \pm 1.9) \times 10^{-6}$ \\
B0218+357 & \citet{mur08} & 0.6847 & $\le 1.8 \times 10^{-6}$ \\
B0218+357 & \citet{kan11} & 0.6847 & $(\le 3.6 \times 10^{-7}$ \\
Milky Way & \citet{lev08} & 0.0 & $\le 3 \times 10^{-8}$ \\
Milky Way & \citet{mol09} & 0.0 & $(4 - 14) \times 10^{-8}$ \\
Milky Way & \citet{lev10} & 0.0 & $(26 \pm 3 \times 10^{-9}$ \\
Milky Way & \citet{lev11} & 0.0 & $ \le 2.8\times 10^{-8}$ \\
\hline
\end{tabular}
\caption{Recent Astronomical $\Delta\mu/\mu$ Measurements}  
\end{minipage}
\end{table*}
\label{lastpage}

\begin{thebibliography}{99}
\bibitem[\protect\citeauthoryear{Avelino et al.}{2006}]{ave06} Avelino, P.P, Martins, C.J.A.P., 
       Nunes, N.J. \& Olive, K.A. 2006, Phys. Rev. D., 74, 083508
\bibitem[\protect\citeauthoryear{Balbi \& Quercellini}{2007}]{bal07} Balbi, A. \& Quercellini, C.
	2007, MNRAS, 382, 1623
\bibitem[\protect\citeauthoryear{Barrow \& Magueijo}{2005}]{bar05} Barrow, J.D. \& Magueijo, J.
	2005, Phys. Rev. D, 72, 043521
\bibitem[\protect\citeauthoryear{Carroll}{1998}]{car98} Carroll, S.M. 1998, Phys. Rev. Let. 81, 3067
\bibitem[\protect\citeauthoryear{Chand et al.}{2004}]{cha04} Chand, H., Srianand, R., Petitjean, P., 
      and Aracil, B.. 2004, A\&A 417, 853
\bibitem[\protect\citeauthoryear{Chiba, Dutta \& Scherrer}{2009}]{chi09} Chiba, T., Dutta, S. \&
	Scherrer, R.J. 2009, Phys. Rev. D., 80, 043517
\bibitem[\protect\citeauthoryear{Coc et al.}{2007}]{coc07} Coc, A., Nunes, N.J., Olive, K.A.,
	Uzan, J-P \& Vangioni, E. 2007, Phys. Rev. D, 76, 023511
\bibitem[\protect\citeauthoryear{Copeland et al.}{2004}]{cop04} Copeland, E.J., Nunes, N.J. \&
        Pospelov, M. 2004, Phys. Rev. D, 69, 023501
\bibitem[\protect\citeauthoryear{Cowie \& Songaila}{1995}]{cow95} Cowie, L.L. \& Songaila, A. 
	1995, Ap.J., 453, 596
\bibitem[\protect\citeauthoryear{Curran et al.}{2011}]{cur11} Curran, S.J. et al. 2011, AAP,
	533, A55
\bibitem[\protect\citeauthoryear{Dutta \& Scherrer}{2008}]{dut08} Dutta, S. \& Scherrer, R.J. 2008,
	Phys. Rev. D., 78, 123525
\bibitem[\protect\citeauthoryear{Dutta \& Scherrer}{2011}]{dut11} Dutta, S. \& Scherrer, R.J. 2011, 
       Physics Letters B, Volume 704, Issue 4, p. 265-269.
\bibitem[\protect\citeauthoryear{Ellingsen, Voronkov, Breen \& Lovell}{2012}]{ell12} Ellingsen, S.P.,
	Voronkov, M.A., Breen, S.L. \& Lovell, E.J. 2012, Ap.J.L, 747, L7.
\bibitem[\protect\citeauthoryear{Flambaum \& Kozlov}{2007}]{fla07} Flambaum, V.V. \& Kozlov, M.G.
	2007, PRL, 98, 240801
\bibitem[\protect\citeauthoryear{Foltz, Chaffee \& Black}{1988}]{fol88} Foltz, C.B., Chaffee, F.H.,
	Black, J.H. 1988, Ap.J., 324, 267
\bibitem[\protect\citeauthoryear{Gupta, Saridakis \& Sen}{2009}]{gup09} Gupta, G., Sridakis, E.N.
	\& Sen, A.A. 2009 Phys. Rev. D, 79, 123013
\bibitem[\protect\citeauthoryear{Henkel et al.}{2009}]{hen09} Henkel, C. et al. 2009, A\&A, 500, 725
\bibitem[\protect\citeauthoryear{Ivanchik, Rodriguez, Petitjean \& Varshalovich}{2002}]{iva02}
	Ivanchik, A.V., Rodriguez, E., Petitjean, P. \& Varshalovich, D.A. 2002, Astronomy
	Letters, 28, 423, (page 483 in Astronomicheskii Zhurnal)
\bibitem[\protect\citeauthoryear{Ivanchik, Petitjean, Rodriguez \& Varshalovich}{2003}]{iva03}
	Ivanchik, A.V., Petitjean, P., Rodriguez, E. \& Varshalovich, D.A. 2003, A\&A Sup., 283, 583
\bibitem[\protect\citeauthoryear{Ivanchik et al.}{2005}]{iva05} Ivanchik, A.V. et al. 2005, A\&A,
	440, 45
\bibitem[\protect\citeauthoryear{Kanekar}{2011}]{kan11} Kanekar, N. 2011, Ap.J.L., 728, L12
\bibitem[\protect\citeauthoryear{King et al.}{2009}]{kin09} King, J. A., Webb, J. K., Murphy, M. T. \&
	Carswell, R. F. 2009, PRL, 101, 251304
\bibitem[\protect\citeauthoryear{King et al.}{2011}]{kin11} King, J. A., Webb, J. K., Murphy, M.,
        Ubachs, W, \& Webb, J. 2011, MNRAS, 417, 3010
\bibitem[\protect\citeauthoryear{King et al.}{2012}]{kin12} King, J. A. et al. 2012, 
	Mon. Not. R. Astron. Soc. 422, 3370
\bibitem[\protect\citeauthoryear{Levshakov, Agafonova, Molaro, \& Reimers}{2008}]{lev08} 
	Levshakov, S.A., Agafonova, I.I., Molaro, P., \& Reimers, D. 2008, Mem. S.A. It. 80, 850
\bibitem[\protect\citeauthoryear{Levshakov et al.}{2010}]{lev10} Levshakov, S.A. et al. 2010, 
	A\&A 524, A32
\bibitem[\protect\citeauthoryear{Levshakov, Kozlov \& Reimers}{2011}]{lev11} Levshakov, S.A., 
	Kozlov, M.G. \& Reimers, D. 2011, Ap.J., 738, 26
\bibitem[\protect\citeauthoryear{Loeb}{1998}]{loe98} Loeb, A. 1998, Ap.J.L., 499, L111
\bibitem[\protect\citeauthoryear{Malec et al.}{2010}]{mal10} Malec, A.L. et al. 2010, MNRAS, 403, 1541
\bibitem[\protect\citeauthoryear{Molaro et al.}{2009}]{mol09} Molaro, P., Levshakov, S.S. 
       \& Kzolov, M.G. 2009, Nuc. Phys. B Proc. Supp., 194, 287
\bibitem[\protect\citeauthoryear{Muller et al.}{2011}]{mul11} Muller, S. et al. 2011, arXiv:1104.3361v1
\bibitem[\protect\citeauthoryear{Murphy et al.}{2004}]{mur04} Murphy, M.T., Webb, J.K. 
            \& Flambaum, V.V. 2004, MNRAS, 345, 609
\bibitem[\protect\citeauthoryear{Murphy et al.}{2008}]{mur08} Murphy, M.T., Flambaum, V.V., Muller, S.,
        \& Henkel, C. 2008, Science, 320, 1611
\bibitem[\protect\citeauthoryear{Nunes \& Lidsey}{2004}]{nun04} Nunes, N.J. \& Lidsey, J.E. 2004, Phys 
      Rev D, 69, 123511
\bibitem[\protect\citeauthoryear{Nunes, Dent, Martins \& Robbers}{2009}]{nun09} Nunes, N.J., Dent, T.,
	Martins, C.J.A.P. \& Robbers, G. 2009, Mm. SAI, 80, 785
\bibitem[\protect\citeauthoryear{Potekhin et al.}{1998}]{pot98} Potekhin, A. Y. et al. 1998, Ap.J.,
	505, 523
\bibitem[\protect\citeauthoryear{Reinhold et al.}{2006}]{rei06} Reinhold, E. et al. 2006, Phys. Rev. 
       Lett., 96, 151101.
\bibitem[\protect\citeauthoryear{Scherrer \& Sen}{2008}]{sch08} Scherrer, R.J. \& Sen, A.A. 2008, 
       Phys. Rev. D, 77, 083515
\bibitem[\protect\citeauthoryear{Thompson}{1975}]{thm75} Thompson, R.I., 1975, Astrophysical Letters, 16, 3
\bibitem[\protect\citeauthoryear{Thompson et al.}{2009}]{thm09} Thompson, R.I. et al. 2009, Ap.J., 
     703, 1648
\bibitem[\protect\citeauthoryear{Thompson}{2012}]{thm12} Thompson, R.I., 2012 MNRAS Letters, 422, L67
\bibitem[\protect\citeauthoryear{Ubachs \& Reinhold}{2004}]{uba04} Ubachs, W. \& Reinhold, E. 2004,
	Phys. Rev. Let., 92, 101302-1
\bibitem[\protect\citeauthoryear{Ubachs et al.}{2007}]{uba07} Ubachs, W., Buning, R., Eikema, K.S.E. 
     \& Reinhold,E., 2007, Journal of Molecular Spectroscopy, 241, 155
\bibitem[\protect\citeauthoryear{Vielzeuf and Martins}{2012}]{vie12} Vielzeuf, P.E. \& Martins, 
	C.J.A.P. 2012, Phys. Rev. D, 85, 087301
\bibitem[\protect\citeauthoryear{Webb et al.}{2011}]{web11} Webb, J.K., King, J.A., Murphy, M.T., 
     Flambaum, V.V., Darswell, R.F., \& Bainbridge, M.B. 2011, PRL, 107, 191101-1-5
\bibitem[\protect\citeauthoryear{Wendt \& Reimers}{2008}]{wen08} Wendt, M. \& Reimers, D. 2008, 
      Eur. Phys. J. ST, 163, 197
\bibitem[\protect\citeauthoryear{Wendt \& Molaro}{2011}]{wen11} Wendt, M. \& Molaro, P. 2011, A\&A,
	526, A96
\bibitem[\protect\citeauthoryear{Wendt \& Molaro}{2012}]{wen12} Wendt, M. \& Molaro, P. 2012, A\&A,
	541, A69
\end{thebibliography}
\end{document}